\documentclass[a4paper,twoside]{article}

\usepackage{epsfig}
\usepackage{subcaption}
\usepackage{calc}
\usepackage{amssymb}
\usepackage{amstext}
\usepackage{amsmath}
\usepackage{amsthm}
\usepackage{multicol}
\usepackage{pslatex}
\usepackage{graphicx}
\usepackage{url}
\usepackage{algorithm2e}
\usepackage[bottom]{footmisc}
\usepackage{multirow}
\usepackage{natbib}
\usepackage{SCITEPRESS}     

\begin{document}

\title{Exploring Robust Intrusion Detection: A Benchmark Study of Feature Transferability in IoT Botnet Attack Detection}

\author{\authorname{Alejandro Guerra-Manzanares\sup{1}\orcidAuthor{0000-0002-3655-5804}\thanks{ \scriptsize{Corresponding author: alejandro.guerra@nottingham.edu.cn}} and Jialin Huang\sup{2}}
\affiliation{\sup{1}School of Computer Science, University of Nottingham, Ningbo, China}
\affiliation{\sup{2}School of Computer Science, University of Nottingham, Nottingham, United Kingdom}
}


\abstract{Cross-domain intrusion detection remains a critical challenge due to significant variability in network traffic characteristics and feature distributions across environments. This study evaluates the transferability of three widely used flow-based feature sets (Argus, Zeek and CICFlowMeter) across four widely used datasets representing heterogeneous IoT and Industrial IoT network conditions. Through extensive experiments, we evaluate in- and cross-domain performance across multiple classification models and analyze feature importance using SHapley Additive exPlanations (SHAP). Our results show that models trained on one domain suffer significant performance degradation when applied to a different target domain, reflecting the sensitivity of IoT intrusion detection systems to distribution shifts. Furthermore, the results evidence that the choice of classification algorithm and feature representations significantly impact transferability. Beyond reporting performance differences and thorough analysis of the transferability of features and features spaces, we provide practical guidelines for feature engineering to improve robustness under domain variability. Our findings suggest that effective intrusion detection requires both high in‑domain performance and resilience to cross‑domain variability, achievable through careful feature space design, appropriate algorithm selection and adaptive strategies.}

\onecolumn \maketitle \normalsize \setcounter{footnote}{0} \vfill

\section{\uppercase{Introduction}}
\label{sec:introduction}

The Internet of Things (IoT) represents a major milestone in the evolution of the global information sector, building upon the foundation established by the Internet. At its core, IoT is an intelligent network that enables devices to exchange information and communicate continuously over the Internet. This connectivity allows individuals to track, monitor, locate, identify and control a wide range of objects and devices with unprecedented efficiency~\citep{alissa2022botnet}. 

The number of connected IoT devices reached 18.5 billion in 2024 and is projected to grow by 14\% in 2025, reaching 39 billion by 2030 and surpassing 50 billion by 2035~\citep{iot_prediction}. This exponential growth brings significant benefits in automation, efficiency and connectivity, but it also introduces critical security challenges. IoT devices often suffer from inherent limitations, including restricted processing power, limited storage capacity and constrained battery life, which hinder the implementation of robust security controls~\citep{rozlomii2024data}. Further, the IoT ecosystem remains in a developmental stage, lacking standardized security models and frequently prioritizing functionality over protection. As a result, IoT networks present numerous exploitable attack vectors~\citep{hussain2020machine}.

A major threat in this context is the rise of botnets. A botnet is a network of \emph{infected} devices that can execute tasks without direct human involvement and be orchestrated into a powerful system capable of executing large-scale automated operations, such as massive spam campaigns and Distributed Denial-of-Service (DDoS) attacks~\citep{report_2025}. The implications of successful IoT botnet attacks extend far beyond the compromise of individual devices. Since IoT systems are integrated into critical infrastructure sectors, such as healthcare, energy and transportation, botnet attacks can result in major service disruptions, data breaches and privacy violations~\citep{mims2025botnet}. IoT devices in homes also constitute a significant security challenge, as their vulnerabilities are often exploited by attackers~\citep{report_2025}

As IoT botnet attacks become more sophisticated, traditional signature- and rule-based detection methods are ineffective against emerging threats due to their limited adaptability to diverse IoT devices and communication protocols, particularly against zero-day or novel attacks~\citep{abraham2018comparison}. To address this, researchers are increasingly using Machine Learning (ML) methods to build intelligent detection systems~\citep{guerra2019hybrid}. However, current approaches face several challenges that limit their effectiveness in real-world deployment scenarios: poor cross-domain generalization, reliance on attack detection, overdependence on specific datasets and limited interpretability~\citep{guerra2019towards}. These limitations reduce effectiveness against zero-day threats and diverse attack patterns arising from variations in device types, network topologies and traffic behaviors.

In addition, feature heterogeneity poses significant challenges for IoT botnet detection, as different feature extraction tools produce incompatible representations. Tools such as Zeek, Argus and CICFlowMeter generate different feature sets with varying dimensionalities and semantic meanings, complicating the development of universally applicable detection systems~\citep{ebrahem2025towards}. These differences can lead to substantial variations in detection performance, emphasizing the fundamental difficulty of achieving cross-tool compatibility in IoT security.

Current work provides only a limited understanding of feature transferability, as most evaluations emphasize in-domain performance rather than cross-domain generalization~\citep{ferrag2020deep}. Early-stage detection, crucial for preventing botnet attacks, has also received insufficient attention~\citep{guerra2020using}. As a result, detection models often lack adaptability to new IoT environments, forcing practitioners to rely on heavily customized solutions that increase deployment costs, prolong implementation and increase risks during adaptation.

A major constraint for model deployment is that deep learning (DL) requires large amounts of training data and computational resources, whereas classical ML methods can achieve strong performance with less data and fewer computational resources for both training and inference. Finally, the lack of systematic feature transferability analysis hinders the development of scalable IoT security solutions, leaving practitioners without clear guidance on which features generalize across different IoT environments.

Addressing these gaps, this study advances IoT botnet attack detection focusing on feature transferability across diverse environments. Our main contributions are:

\begin{enumerate}
    \item \textbf{Benchmarking feature transferability}: We implement a unified framework to evaluate the transferability of widely used feature extraction tools across multiple IoT datasets, establishing a standard reference beyond single‑dataset evaluations. Models are trained and tested on features from the same tool, ensuring fair comparison and eliminating forced feature alignment.

    \item \textbf{Focus on cross-domain model evaluation}: We evaluate four widely used ML algorithms under both in‑ and cross‑domain settings. In‑domain tests measure performance when training and testing use the same dataset, while cross‑domain tests apply models under zero‑adaptation conditions to new datasets, simulating real‑world deployment and highlighting performance under distribution shift.

   \item \textbf{Practical reference for intrusion detection design}: Our results and analysis provide relevant insights to help researchers and practitioners design scalable intrusion detection systems for diverse IoT environments.

\end{enumerate}

This paper is organized as follows: Section~\ref{related_work} reviews related literature, Section~\ref{method} outlines the benchmark methodology and Section~\ref{results} presents the main findings. Section~\ref{discussion} discusses the results, while Section~\ref{conclusion} concludes the study.

\section{\uppercase{Related Work}}
\label{related_work}

Research has extensively explored ML and DL approaches to address IoT network security challenges, particularly in botnet detection and intrusion prevention. \citet{bertino2017botnets} highlighted the limitations of conventional testing methods in heterogeneous and rapidly evolving IoT environments, where zero-day attacks remain difficult to counter. Based on this, \citet{disha2022performance} proposed a Gini index weighted random forest that demonstrated superior performance on UNSW-NB15 and TON IoT datasets, while~\citet{gu2021effective} combined Naive Bayes with SVM to achieve high accuracy across multiple datasets. \citet{sinha2025high}. demonstrated the robustness of CNN-LSTM hybrids on the BoT-IoT dataset, reaching 99.87\% accuracy while maintaining resilience under noise through SHAP-based feature selection and SMOTE balancing. Lightweight architectures such as SkipGateNet achieved 99.91\% accuracy with minimal storage and rapid processing, enabling deployment on constrained IoT devices~\citep{alshehri2024skipgatenet}. \citet{abbas2024evaluating} emphasized the need for adaptable DL models across diverse operational contexts. Despite these advances, challenges persist due to the lack of standardization and the highly variable nature of IoT data, which hinder model generalization across diverse environments.

Transfer learning (TL) has emerged as a crucial approach in IoT security, enabling knowledge reuse across domains with limited labeled data and heterogeneous environments. \citet{rodriguez2022transfer} demonstrated its effectiveness by integrating CNN with TL to improve zero-day attack detection through feature freezing and unsupervised adaptation. However, \citet{li2023maximal} found that domain mismatch and distributional shifts often hinder generalization of pretrained models. Overall, TL reduces distributional gaps and boosts detection accuracy, though future work must separate feature transferability from algorithmic adaptation to achieve robust generalization in dynamic IoT settings.

\citet{nazir2023advancing} highlighted inconsistencies in datasets, feature extraction and metrics, emphasizing the need for unified protocols. Although datasets like BoT-IoT~\citep{koroniotis2019towards} are valuable benchmarks covering multiple attack types, concerns arise about their ability to reflect real-world traffic. To address these issues, \citet{keshk2023explainable} introduced a framework combining explainable DL with cross-domain evaluation, enhancing transparency and reliability. \citet{lypa2025comparison} compared traffic analysis tools showing differences in generated features and resource demands. Their study highlights the need to select appropriate tools for IoT security, as feature extraction directly impacts model transferability.

Despite progress in IoT botnet detection, significant gaps remain regarding feature transferability. Most studies focus on single datasets, achieving high in-domain accuracy but neglecting cross-dataset or cross-environment evaluations, leaving robustness and generalization unaddressed. Research also lacks systematic comparisons across widely used tools and little attention has been given to identifying which feature categories retain stability across domains. Current work relies on tool-specific benchmarks without clarifying whether extracted features can generalize beyond their original environment. To address these limitations, our work establishes a structured benchmark using popular tools across multiple IoT botnet datasets under zero-adaptation conditions, providing a systematic evaluation of feature transferability in heterogeneous IoT settings.

\section{\uppercase{Methodology}}
\label{method}

This section outlines the benchmarking framework implemented in our study, including the procedures used to evaluate feature transferability between domains. It also details the experimental setup, the datasets employed, the feature spaces, in addition to the ML classification algorithms and performance metrics applied.

\begin{figure*}[t]
    \centering
    \includegraphics[width=\linewidth]{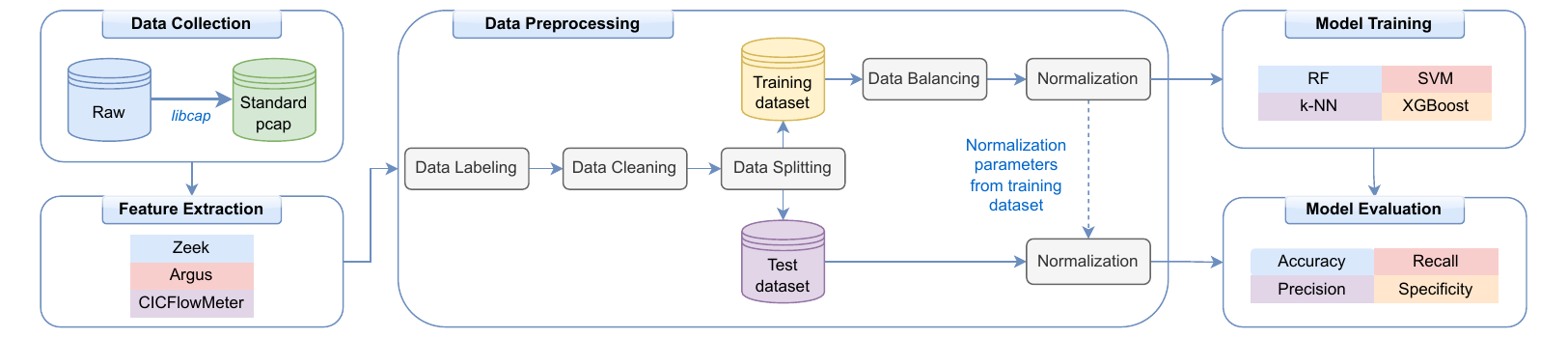}
    \caption{Benchmarking framework overview}
    \label{fig:benchmarking}
\end{figure*}

\subsection{Benchmarking Framework}

This study implements a benchmarking framework to evaluate how different feature extraction methods transfer knowledge between domains in IoT botnet detection systems. As depicted in Figure~\ref{fig:benchmarking}, the framework consists of five main steps:

\begin{enumerate}
    \item \textbf{Data Collection:} Four representative public datasets (MedBIoT, TON IoT, Edge-IIoTset, CICIoT2023) are selected, covering both general IoT and industrial IoT application scenarios to ensure diversity of the experimental setup.

    \item \textbf{Feature Extraction:} Three widely used feature extraction tools (Zeek, Argus and CICFlowMeter) are employed to transform raw traffic packets (\emph{pcap} files) into structured feature representations. These tools provide different levels of granularity and semantic content, enabling systematic cross-domain comparisons.

    \item \textbf{Data Preprocessing:} Includes uniform standardization of all feature sets through label encoding, missing value imputation, outlier truncation, normalization and class balancing. These steps are designed to remove noise arising from heterogeneous data sources and ensure comparability across domains.

    \item \textbf{Model Training:} Classic ML algorithms (Random Forest (RF), Support Vector Machines (SVM), k-Nearest Neighbors (k-NN) and XGBoost) are trained on source domain datasets and tested on target domain data without retraining. This simulates realistic deployments where models must generalize to unseen environments or adapt to drastic changes in operating conditions.

    \item \textbf{Evaluation:} Cross-domain detection performance is systematically assessed using relevant performance metrics, i.e., accuracy, recall, precision and specificity. To control for randomness bias, we perform repeated experiments and report confidence intervals.
\end{enumerate}

This framework enables rigorous assessment of feature extraction tools in cross-domain transfer operations while preserving their original feature semantics. The results provide practical design recommendations for IoT intrusion detection systems.

\subsection{Experimental Setup}

Our benchmark implements a single-source zero-adaptation cross-domain testing strategy, focusing on evaluating the transferability of features generated by different feature extraction tools in cross-domain detection. The experimental process is as follows: each time, one dataset is fixed as the source domain for model training and validation; the remaining three datasets serve as target domains, used only for testing and not for retraining.

For all experiments, the classification threshold is fixed at $0.5$ (the neutral midpoint for binary classification) and no hyperparameter tuning or threshold adjustment is performed on the target domain. This strict setting ensures that there is no information leakage from the target domain, thus providing a fair and rigorous evaluation of cross-domain generalization ability. Although this setting may lead to extreme cases, such as perfect scores when source and target domains are highly similar, or high recall but low precision when threshold misalignment occurs, the results are considered normal and interpretable within the transfer learning evaluation framework.

All datasets are processed using three feature extraction tools (Zeek, Argus, CICFlowMeter) to form three distinct feature spaces. Within each feature space, four classifiers (RF, SVM, k-NN, XGBoost) are selected for model building. To reduce the influence of random factors, each experiment is repeated five times and the average results are reported.

To properly assess cross-domain performance, we also conduct in-domain evaluation, which measures the model’s performance when the source and target domains are the same and serves as a reference baseline for evaluating cross-domain results.

\subsection{Datasets}

To establish a diverse framework for evaluating feature transferability across domains, we use four publicly available IoT botnet datasets. Detailed descriptions of these datasets are provided as follows. 
 Note that all datasets include benign or normal IoT operation data and simulated malicious behaviors.

\begin{itemize}
    \item \textbf{MedBIoT}~\citep{guerra2020medbiot}. It provides network traffic from a medium-scale realistic IoT network comprising 83 physical and emulated devices, distributed across multiple subnets to emulate office environments. Using real malware samples (i.e., Mirai, Bashlite, Torii), the dataset focuses on capturing botnet propagation activity rather than attack execution.
    \item \textbf{CICIoT2023}~\citep{neto2023ciciot2023}. It is a large-scale IoT dataset, including 105 real devices, that emulate a smart home environment. It focuses on attack detection, simulating 33 distinct attacks that encompass both conventional threats (e.g., web-based and brute-force attacks) and IoT-specific attacks (e.g., DoS, DDoS and Mirai).
    \item \textbf{TON\_IoT}~\citep{moustafa2021new}. It enables security research in both IoT and IIoT domains by providing diverse data sources such as telemetry from IoT and IIoT sensors, Windows-based systems and Linux servers. Using a realistic virtualized topology, it contains network traffic from real and emulated smart home devices, edge nodes and industrial sensors, focusing on attack simulation.
    \item \textbf{Edge-IIoTset}~\citep{ferrag2022edge}. It allows research in industrial and edge computing environments by providing network traffic from seven domains such as agriculture, healthcare, manufacturing and autonomous vehicles. It focuses on simulating 14 distinct attacks within a topology composed of 13 real IoT and IIoT sensors.
\end{itemize}

\subsection{Feature Spaces}

In this study, we use three feature extraction tools to derive statistical representations (i.e., features) from raw network traffic. Their selection is motivated by their widespread adoption in ML-based intrusion detection research and their ability to generate distinct feature spaces that effectively capture the behavioral dynamics of IoT environments. The extraction tools are described as follows. 

\begin{itemize}
    \item \textbf{CICFlowMeter}~\citep{cicflowmeter}. Formerly known as ISCXFlowMeter, it is a Java-based network traffic analysis tool that constructs bidirectional flows using a standard 5-tuple (source IP, destination IP, source port, destination port and protocol) and extracts statistical features per flow, capturing traffic behavior across both directions through several metrics, time- and size-related attributes.
    \item \textbf{Zeek}~\citep{zeek}. Formerly known as Bro, it is an open-source network monitoring framework designed as a passive traffic analyzer. Rather than focusing on packet-level statistics, it interprets protocol semantics and generates structured logs that capture session or connection-level details such as duration, byte counts, TCP flags and state transitions. 
    \item \textbf{Argus}~\citep{argus}. Short for Audit Record Generation and Utilization System (ARGUS), it is an open-source network flow monitoring tool designed for continuous traffic analysis. It generates bidirectional flow records and reports detailed attributes that cover network identification, packet dynamics, services and resource utilization.
\end{itemize}

To evaluate feature transferability, the original \emph{pcap} traffic files from each dataset were processed using the three feature extraction tools. Although applied to the same input data, the tools produce outputs that vary in structure, type and feature count, producing three distinct feature sets.

Note that before feature extraction, all traffic captures were standardized to the \emph{libpcap} format (\emph{.pcap}) to ensure tool compatibility and address format inconsistencies. This established consistent processing conditions, ensuring fairness, reproducibility and reliability in the transferability analysis.

\subsection{ML Workflow}

The following sections describe the data preprocessing steps applied to prepare the datasets, the ML models evaluated and the performance metrics assessed.

\subsubsection{Data Preprocessing}

\textbf{Label Encoding}. Traffic labels are consolidated into a binary scheme to ensure consistency across datasets. Specifically, all benign flows are assigned a value of 0, while all attack categories are grouped into a single class with a value of 1. This eliminates issues arising from different labeling schemes and number of classes across datasets. 

\textbf{Data Cleaning}. The datasets used in this study originate from heterogeneous sources and experimental configurations, which inevitably introduce inconsistencies in structure, content and feature definitions. These discrepancies can introduce noise and bias, undermining model performance and reducing feature transferability. The data cleaning process comprises three main stages: (i) removal of environment-specific identifiers such as source and destination IP addresses, ports and timestamps, which encode deployment-specific characteristics that may lead to overfitting in transfer learning; (ii) exclusion of instances with incomplete or corrupted feature values, especially when multi-tool extraction introduces segmentation mismatches; (iii) imputing missing numerical values using the median value for that feature, or zero if that feature is entirely missing on the dataset; and (iv) standardization of feature names and column ordering across extraction tools to ensure dimensional and semantic alignment between datasets.


\textbf{Data Splitting}. The final dataset was constructed in 12 combinations derived from the cross-product of the three feature extraction tools and the four IoT botnet datasets. For each combination, samples were split hierarchically into training and test sets with an 8:2 ratio. To reflect realistic evaluation conditions, the test sets preserved the original class distributions, while balancing techniques were applied only to the training sets. This ensures that the results report actual cross-domain generalization performance without introducing artificial bias into the test stage.

\textbf{Data Balancing}. A major challenge is the extreme class imbalance present in certain datasets, particularly TON\_IoT, where benign traffic is scarce and malicious traffic is dominated by DoS and scanning events. In contrast, attack types such as ransomware are rare. This skewed distribution biases classifiers toward majority classes, often yielding high overall accuracy but poor recall on minority categories. To mitigate this issue, we used the Synthetic Minority Over-sampling Technique for Nominal and Continuous features (SMOTE-NC)~\citep{chawla2002smote}.

\textbf{Normalization}. The datasets contain attributes with diverse numerical ranges, where some variables show significantly larger magnitudes than others. Given that some algorithms are sensitive to extreme values, this may introduce bias towards features with extreme values. To address this, normalization was applied to rescale each feature into the $[0, 1]$ range. Normalization parameters are derived only from the training set and subsequently applied to both training and test sets, ensuring a realistic evaluation scenario and preventing data leakage.

\subsubsection{ML Algorithms}

In this study, four widely used ML algorithms were employed. These algorithms have demonstrated strong performance across diverse domains and are often adopted as baselines in cybersecurity research due to their robustness, interpretability and efficiency. All algorithms were evaluated using typical hyperparameters to ensure fair and comparable results. The selected models are: Random Forest (RF), Support Vector Machines (SVM), $k-$Nearest Neighbors ($k-$NN) and eXtreme Gradient Boosting (XGBoost). The RF classifier was constructed with 300 estimators, ensuring sufficient ensemble diversity for stable performance. The k-NN classifier was configured with $k=5$, using the Euclidean distance metric to measure similarity. XGBoost was trained with 400 boosting rounds, balancing computational efficiency with predictive accuracy. The SVM classifier was implemented with a linear kernel, enabling efficient handling of high-dimensional feature spaces.

\subsubsection{Performance metrics}

The evaluation is conducted using four relevant performance metrics derived from the confusion matrix: accuracy, recall, precision and specificity. \emph{Accuracy} measures the overall proportion of correctly classified instances (both positive and negative) out of all predictions, providing a broad view of model performance. However, in highly imbalanced datasets, accuracy can be misleading, as a model may achieve high accuracy simply by favoring the majority class while neglecting minority instances. In our case, we ensured the reliability of the accuracy metric by using balanced training, validation and test sets. \emph{Recall}, also known as sensitivity or True Positive Rate (TPR), reflects the ability of the model to correctly identify positive instances, making it particularly important in scenarios where missing minority classes (e.g., rare attack types) would be costly. \emph{Precision}, on the other hand, reflects the correctness of positive predictions, penalizing false alarms and ensuring that detected events are truly relevant. Finally, \emph{Specificity}, also known as True Negative Rate (TNR), reports the model’s ability to correctly identify negative (normal) instances. It is particularly important in contexts where false positives are costly, ensuring that normal data is not incorrectly flagged as positive.


\section{\uppercase{Results}}
\label{results}

The following section reports the results of the feature transferability analysis in- and cross-domain across classification models and datasets. It also provides an analysis of the most important features of each feature space for in- and cross-domain detection models.

\subsection{Feature Set Transferability Across Classification Models}
\label{sec:transferability_models}

\begin{figure*}[!h]
  \centering
   {\includegraphics[width=\linewidth]{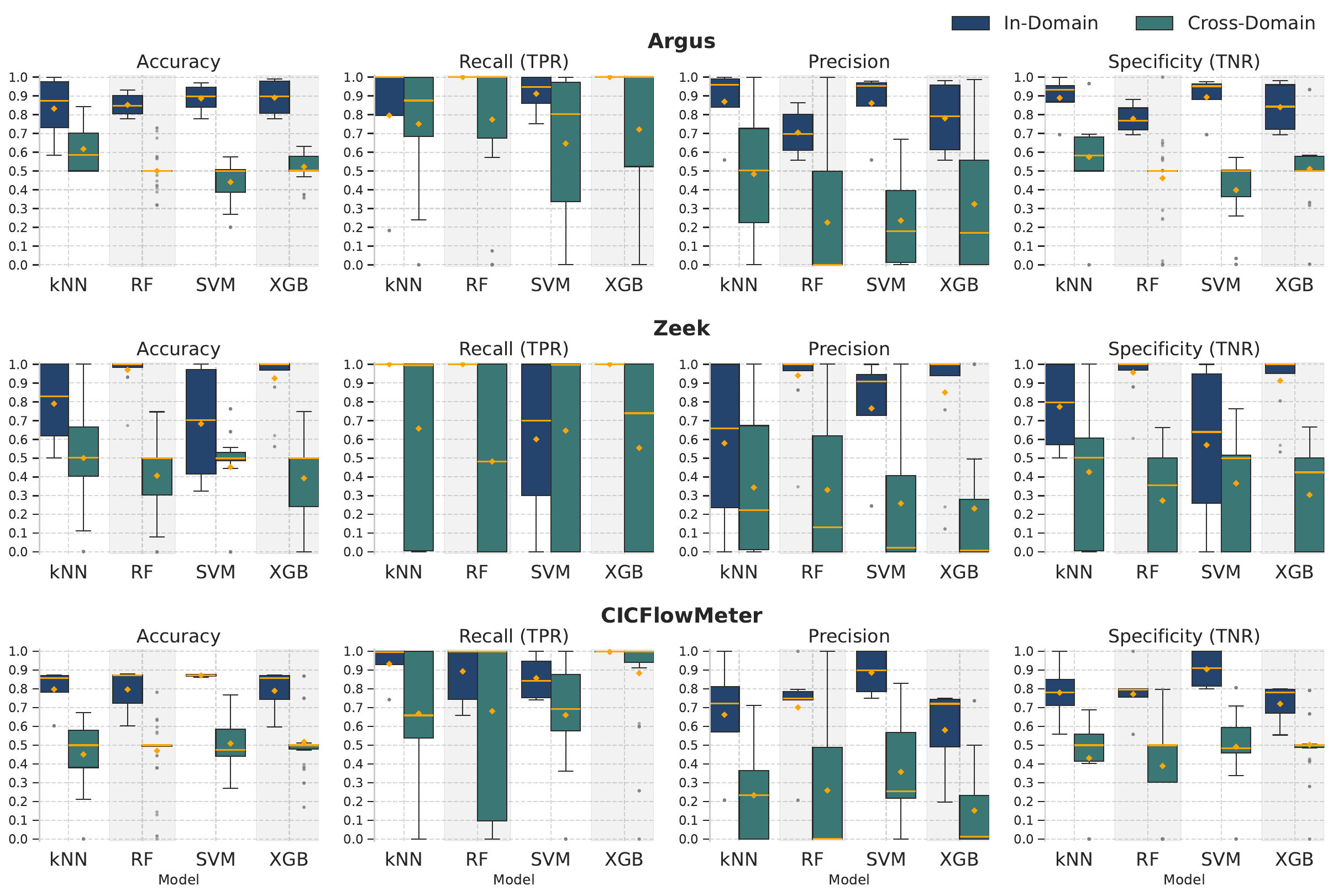}}
  \caption{In- and cross-domain performance of feature spaces across classification models}
  \label{fig:classification_models}
 \end{figure*}

Figure~\ref{fig:classification_models} provides a comparative visualization of the performance metrics obtained for the four classification algorithms in both in- and cross-domain evaluation settings. The figure uses boxplots to capture the variability in the results, providing a concise yet informative summary of the distribution of performance scores. Each boxplot highlights the median and average values as measures of central tendency, using an orange horizontal line in the body of the boxplot and an orange rhombus, respectively. The interquartile range (IQR), delimited by the boxplot body, reports the spread and consistency of the models’ performance. Outliers are also depicted, reporting cases in which the algorithms showed unusually high or low accuracy relative to the bulk of the results.

By juxtaposing in- and cross-domain outcomes, Figure~\ref{fig:classification_models} enables a direct assessment of how well each algorithm generalizes beyond the training domain. This comparison is crucial for understanding the robustness of the models and the transferability of the feature spaces, as strong in-domain performance does not necessarily translate to effective cross-domain transferability. 

\subsubsection{In-Domain Performance}

In the in-domain evaluation (blue boxplots in Figure~\ref{fig:classification_models}), all classifiers demonstrate consistently strong performance while revealing notable differences in distribution across feature spaces. Specifically, Argus and CICFlowMeter report similar behavior, with accuracy values of $\approx 0.9$ across all classification models, high recall (above $0.9$, except for SVM) and relatively high precision and specificity metrics ranging between $0.8$ and $0.9$. In contrast, the Zeek feature space shows mixed behavior: while RF and XGB achieve near-perfect accuracy, recall, precision and specificity, kNN and SVM report values below $0.8$ for all metrics. This emphasizes that the choice of feature space has a substantial impact on classifier performance. In particular, while Argus and CICFlowMeter provide stable and balanced results across learning algorithms, Zeek appears more sensitive to the underlying classification model, producing excellent results for tree-based models, but weaker results for distance-based and margin-based approaches. These findings highlight the importance of aligning feature representations with the strengths of specific learning algorithms. 

\subsubsection{Cross-Domain Performance}

Cross-domain evaluation (green boxplots in Figure~\ref{fig:classification_models}) shows consistent behavior across feature spaces, but with significantly different results compared to the in-domain case. In general, for all feature spaces and classifier combinations, the accuracy boxplots show larger bodies, with mean and median values clustering around $0.5$, indicating poor generalization to cross-domain settings. Although some exceptions can be observed, such as the combination of kNN and the Argus feature space, this general trend extends to specificity. Most models evidence high recall, at the expense of very low precision scores, highlighting a tendency to over-predict positive instances. This imbalance suggests that classifiers trained in one domain are prone to misclassifying benign traffic as malicious when applied to unseen domains, inflating recall while severely compromising precision. This behavior is problematic in practical deployments, as it can lead to excessive false alarms and undermine the usability of intrusion detection systems. 

It is worth noting that while the above trends apply to all feature spaces, the Argus feature space reports slightly better and more consistent results for generalization, whereas the Zeek feature space performs notably worse than CICFlowMeter and Argus, particularly in terms of recall. The extremely wider bodies observed in Zeek’s boxplots indicate instability and high variability across classifiers, suggesting that its feature representations may be less robust when transferred to cross-domain settings. This instability highlights the susceptibility of certain feature extraction methods to domain shifts, where differences in traffic characteristics or protocol-level details can significantly affect classifier reliability. In contrast, the relatively stable performance of Argus suggests that its features capture slightly more generalizable patterns, making them less sensitive to distributional changes. 

\subsection{Feature Set Transferability Across Domains}

\begin{figure*}[!h]
  \centering
   {\includegraphics[width=\linewidth]{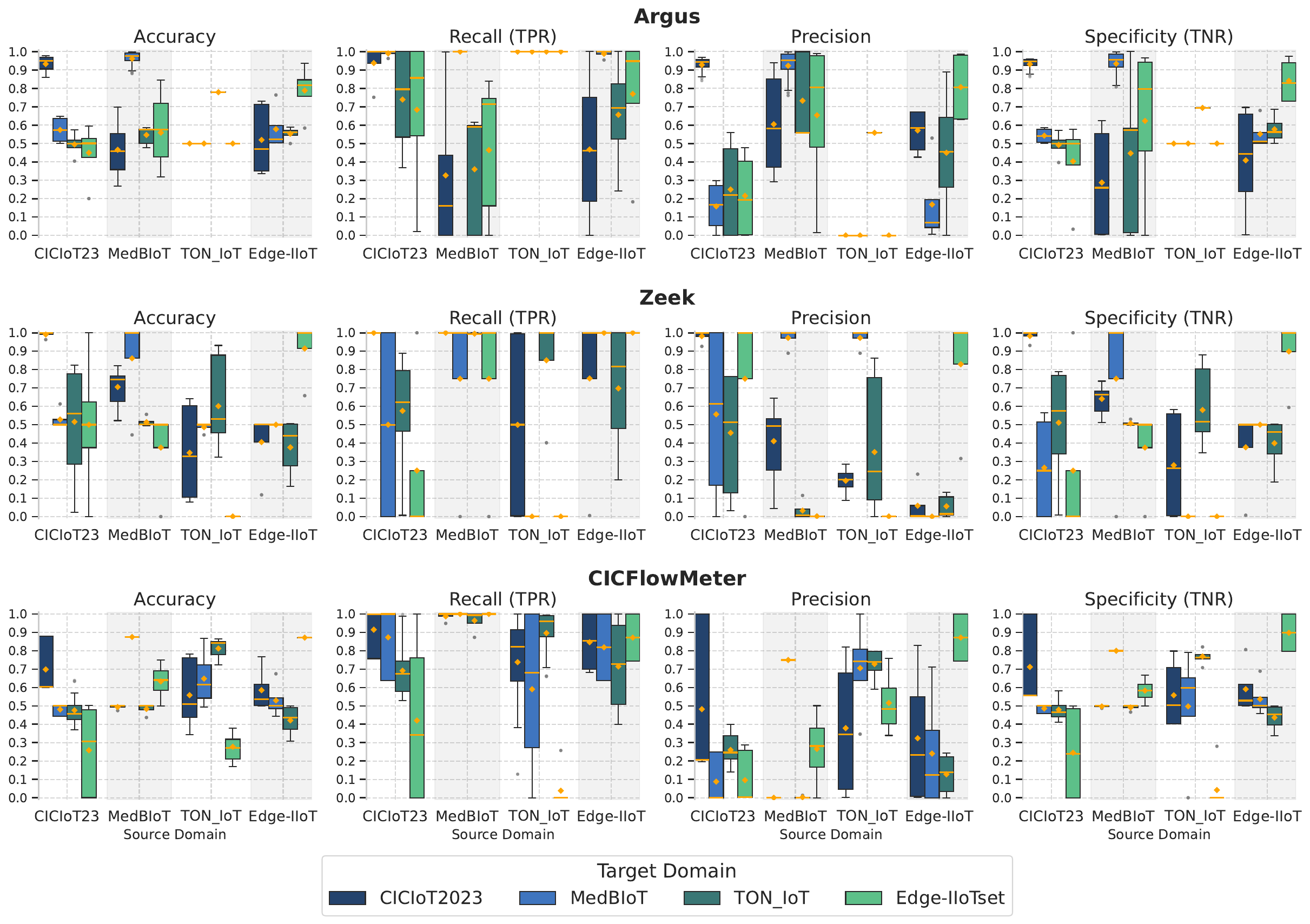}}
  \caption{In- and cross-domain performance of feature spaces across source datasets}
  \label{fig:cross_domain}
 \end{figure*}

The boxplots in Figure~\ref{fig:cross_domain} provide a comparative visualization of performance metrics across four datasets under both in- and cross-domain evaluation settings. In this figure, the x-axis denotes the source domain (i.e., the dataset used for training), while the y-axis represents the generalization performance for in- and cross-domain scenarios, as indicated in the legend (target domain). To further enrich the comparison, color coding is applied in Figure~\ref{fig:cross_domain}: blue shades (dark and light) represent IoT datasets (general IoT domain), while green shades (dark and light) correspond to IIoT datasets (specific IIoT domain). This comparison provides a complementary perspective to Figure~\ref{fig:classification_models}, emphasizing the ability of models trained on a specific source domain to generalize beyond their original training environment. Such generalization is a critical requirement for real-world IoT and IIoT deployments, where attack patterns and data distributions often vary significantly across domains. Evaluating cross-domain performance not only reveals the robustness and adaptability of the models but also helps identify potential domain-specific overfitting. These insights are essential for guiding future improvements in model design, feature representation and training strategies aimed at achieving reliable security in heterogeneous IoT ecosystems.

\subsubsection{In-Domain Performance}

Figure~\ref{fig:cross_domain} illustrates that when the source and target domains are identical (in-domain), the models achieve consistently high performance metrics. However, when evaluated across different feature spaces, the results vary significantly. Specifically, the CICFlowMeter feature space, which is larger and more complex, shows greater variability and lower mean and median values across all datasets, suggesting potential challenges in feature stability and model generalization. In contrast, the Argus and Zeek feature spaces yield remarkably better in-domain performance. In general, both feature spaces report low variability (shorter boxplot bodies) and superior central tendency measures. These observations indicate that feature space selection plays a critical role in both stability and predictive accuracy, with smaller, well-structured and less redundant feature sets, containing fewer but informative features, yield more reliable outcomes.

\subsubsection{Cross-Domain Performance}

Cross-domain evaluation, reported in Figure~\ref{fig:cross_domain} as boxplots derived from different source and target domain combinations, shows poor generalization across all datasets, regardless of whether the domains belong to IoT or IIoT. These results emphasize the inherent challenge of transferring knowledge between heterogeneous environments where feature distributions and attack patterns differ significantly. Despite this general trend, a few interesting observations emerge. 

Regarding feature space transferability, the Argus feature space produces the most consistent results for accuracy and specificity, as indicated by shorter boxplots, but shows greater inconsistency for recall and precision, particularly for the latter. In contrast, the Zeek feature space shows higher variability in accuracy and specificity but achieves better precision and recall in cross-domain settings, suggesting stronger adaptability to detect diverse attack patterns. Finally, the CICFlowMeter feature space demonstrates the highest variability across datasets for recall and precision, while maintaining relatively compact boxplots for specificity, indicating that its larger feature set may introduce noise affecting certain performance metrics.

Regarding datasets, while overall performance metrics across domains are notably poor, certain source domain and feature space combinations yield remarkable results. For example, the TON\_IoT dataset provides the best cross-domain performance when paired with the CICFlowMeter feature space, suggesting that specific combinations of source data and feature representation can partially mitigate domain shift. Similarly, the MedBIoT dataset demonstrates relatively strong cross-domain performance when evaluated on the CICIoT23 dataset (both belonging to the IoT domain) using the Zeek feature space. These observations reinforce the importance of analyzing feature-domain interactions, as certain combinations seem to mitigate domain shift more effectively than others. Understanding these relationships is crucial for designing models that can generalize across heterogeneous IoT and IIoT environments, where variability in traffic patterns and attack behaviors poses significant challenges for intrusion detection systems.

Finally, it is worth noting that while accuracy and specificity results generally indicate poor cross-domain transferability for most feature spaces, recall is particularly high in some cross-domain settings, often paired with extremely low precision. This imbalance suggests that models tend to over-predict positive cases in unfamiliar domains, capturing most true positives (attacks), but at the cost of a large number of false positives. This behavior can be problematic in IDS, as it may lead to excessive false alarms, significantly compromising operational efficiency. These findings highlight the need for strategies that balance recall and precision, such as threshold calibration or domain adaptation techniques, to achieve reliable performance in heterogeneous environments.


\subsection{Feature Importance}

\begin{figure*}[!ht]
  \centering
   {\includegraphics[width=\linewidth]{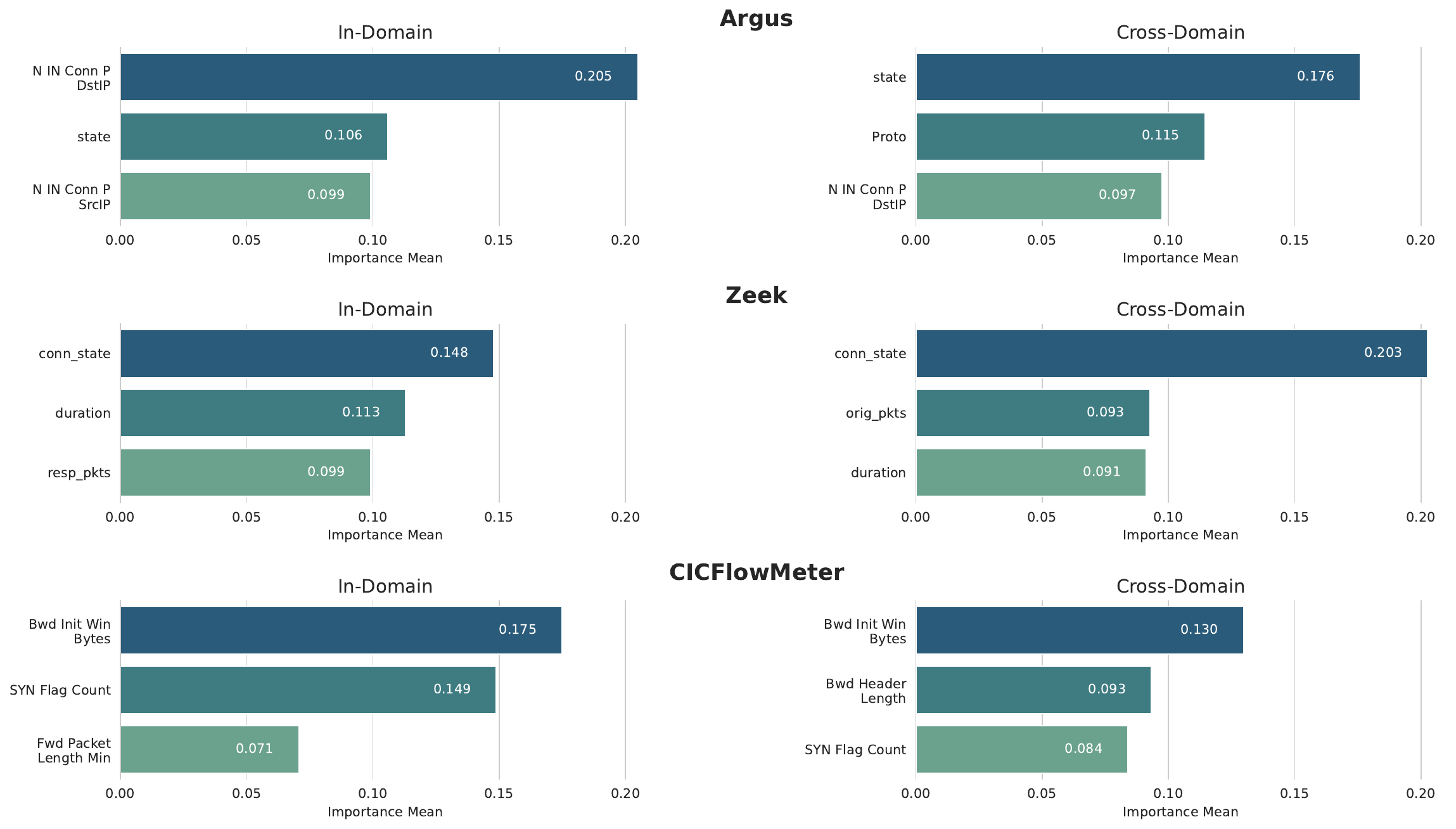}}
  \caption{Top-3 features ranked by importance, based on SHAP values, for in- and cross-domain performance}
  \label{fig:feature_importance}
 \end{figure*}

Since the k-NN model demonstrated slightly better performance in Section~\ref{sec:transferability_models}, we adopt this model to analyze feature importance for both in- and cross-domain settings. To achieve this, we employ SHapley Additive exPlanations (SHAP)~\citep{shap_nips2017} to approximate feature contributions. For each evaluation scenario, SHAP values are computed on the target test set and the mean absolute contribution of each feature is taken as its importance. Features are then ranked according to these values. The in-domain SHAP analysis measures the relative influence of features when models are trained and evaluated within the same domain, whereas the cross-domain SHAP analysis assesses feature influence when models are trained on a source dataset and tested on a different target dataset. The top three features for both in- and cross-domain performance within each feature space are provided in Figure~\ref{fig:feature_importance}, highlighting numerical differences and providing insights into feature relevance under both stable and domain-shift conditions. It is worth noting that this analysis provides a global assessment of feature importance, although the significance of individual features may vary across specific contexts (i.e., different combinations of feature sets, classification models and source–target domain pairs).

Interestingly, across all feature sets, two of the three most influential features remain consistent between in- and cross-domain settings. However, their relative importance and ranking differ significantly across these conditions.

In the Argus feature set, cross-domain analysis ranks \texttt{state} (0.176), \texttt{Proto} (0.115) and \texttt{N IN Conn P DstIP} (0.097) as the top three features. Under in-domain conditions, \texttt{N IN Conn P DstIP} (0.205) becomes the most important feature, nearly doubling its cross-domain value, while \texttt{state} shifts to second (0.115) and \texttt{N IN Conn P SrcIP} (0.099) replaces \texttt{Proto} in third position. Although topology-dependent features (i.e., IP addresses) were removed, Argus still shows a strong reliance on behavioral indicators, such as counts of new connections, which gain prominence under in-domain conditions.

The connection \texttt{state} feature, which represents the lifecycle of network sessions, shows high importance in cross-domain scenarios, almost twice its relevance compared to in-domain. This suggests that session behavior generalizes well across different environments, making it a robust indicator to detect anomalies when training and testing occur in different domains. In contrast, the protocol feature emerges (\texttt{Proto}) as significant only in cross-domain settings, likely because protocol distribution patterns differ between domains.

For the Zeek feature set, the three most influential features for in-domain performance are \texttt{conn\_state}, \texttt{duration} and \texttt{resp\_pkts}, all with importance values between 0.148 and 0.099. For cross-domain, \texttt{conn\_state} remains the most important feature, increasing its weight to 0.203, while \texttt{duration} slightly decreases in importance and \texttt{orig\_pkts} replaces \texttt{resp\_pkts} as the third most relevant feature.

This shift highlights several insights. First, the dominance of \texttt{conn\_state} across both scenarios indicates that session lifecycle behavior generalizes well across domains, making it a robust feature for detecting anomalies in heterogeneous environments. Second, the reduced importance of duration in cross-domain settings suggests that timing characteristics may be more domain-specific, limiting their transferability. Finally, the substitution of \texttt{orig\_pkts} for \texttt{resp\_pkts} under cross-domain conditions reflects a change in traffic directionality patterns between domains, which can serve as a discriminative signal when adapting models to new environments. 

Overall, the findings in the Argus and Zeek important features suggest the need to prioritize behavioral features that capture protocol state and exchange patterns for robust cross-domain intrusion detection.

For the CICFlowMeter feature set, the top three features under cross-domain conditions are \texttt{Bwd Init Win Bytes} (0.130), \texttt{Bwd Header Length} (0.093) and \texttt{SYN Flag Count} (0.084). In in-domain settings, \texttt{Bwd Init Win Bytes} remains the most influential feature, increasing its importance to 0.175. \texttt{SYN Flag Count} also rises significantly to 0.149, while \texttt{Fwd Packet Length Min} (0.073) replaces \texttt{Bwd Header Length} as the third most relevant feature.

This pattern suggests several insights. First, the consistent dominance of \texttt{Bwd Init Win Bytes} across both scenarios suggests that TCP window size behavior is a strong and generalizable indicator of traffic characteristics, making it valuable for both in-domain and cross-domain detection. Second, the increased importance of \texttt{SYN Flag Count} in in-domain conditions indicates that connection initiation patterns become more discriminative when the model is trained and tested within the same environment, likely due to localized attack behaviors such as SYN flooding. Finally, the substitution of \texttt{Fwd Packet Length Min} for \texttt{Bwd Header Length} under in-domain conditions reflects a shift toward finer-grained packet size characteristics, which capture subtle variations in traffic flows that are more relevant within a single domain. Overall, these findings highlight the role of transport-layer and packet-level behavioral features in achieving robust intrusion detection, with TCP-specific indicators being particularly important.

In general, CICFlowMeter demonstrates weaker cross-domain performance primarily due to the nature of its feature set, which is dominated by transport-layer and packet-level indicators such as \texttt{Bwd Init Win Bytes}, \texttt{SYN Flag Count} and header lengths. These features are highly sensitive to protocol configurations and traffic engineering, making them less robust under domain shift. In contrast, Argus and Zeek focus on session-level and behavioral features, such as \texttt{state} and \texttt{conn\_state}, which capture the lifecycle of network connections and protocol states. These abstractions seem to generalize well across heterogeneous environments, explaining their stability and similar top-ranked features in both in- and cross-domain analyzes. The volatility of CICFlowMeter’s packet-centric features, combined with its lack of higher-level behavioral indicators, limits its adaptability when models are transferred between domains. Therefore, Argus and Zeek may provide a more resilient approach to cross-domain intrusion detection by emphasizing protocol state and connection behavior rather than domain-specific packet metrics.





\section{\uppercase{Discussion}}
\label{discussion}

Our experimental results show the inherent challenges of cross-domain feature set transferability: models trained on one domain demonstrate poor generalization when evaluated on a different target domain. Furthermore, the choice of classification algorithm significantly affects cross-domain performance, emphasizing the interplay between features and model architecture in achieving robust transferability.

Beyond algorithmic and source-target domain considerations, our findings show that the nature of the features significantly impacts transferability and their relative importance under in- and cross-domain conditions. Feature space selection is therefore critical for intrusion detection, as it directly influences whether models can maintain performance under domain shift. Our experiments indicate that the Argus and Zeek provide better stability by leveraging behavioral and session-level indicators, while CICFlowMeter's reliance on packet-centric features makes it more vulnerable to distributional changes. This stresses the broader challenge of transferring learned decision boundaries across heterogeneous environments, where differences in traffic characteristics introduce significant variability. Effectively addressing this issue requires strategies such as domain adaptation or designing universal feature representations that capture generalizable patterns. 

The observed differences suggest that, even within in-domain settings, feature extraction strategies can introduce biases that affect generalization. Consequently, robust intrusion detection requires not only careful model selection, but also consideration of the feature space, as mismatches between representation and classifier can lead to significant performance degradation. This implies that effective network security solutions must integrate both algorithmic and feature-engineering perspectives to ensure resilience across diverse traffic patterns and attack scenarios. In conclusion, effective intrusion detection requires not only strong in-domain accuracy but also resilience to cross-domain variability, as real-world deployments inevitably face diverse traffic patterns and evolving attack landscapes.

This study has several limitations: (i) the evaluation was conducted on four IoT botnet datasets, which may not fully represent the heterogeneity of real-world traffic; (ii) zero-adaptation transferability provides a strict benchmark, but does not account for lightweight adaptation techniques that could improve performance; (iii) only three tools were included, which may limit generalization; (iv) feature semantics across tools were not aligned, which could have introduced bias. Despite its limitations, this study offers relevant insights into feature transferability for IoT botnet detection and provides a foundation for future adaptive and scalable network security approaches.

\section{\uppercase{Conclusion}}
\label{conclusion}

Our results confirm the inherent challenges of cross-domain intrusion detection: models trained on one domain demonstrate poor generalization when evaluated on another, with both classification model choice and feature representation significantly influencing transferability. Besides, the nature of the features used is critical to build more resilient models to domain shift. In this regard, we provide guidelines for effective feature engineering based on our experimental results. However, our findings stress the need for strategies such as domain adaptation, feature normalization and the design of universal feature representations to improve robustness under domain variability. Future work will explore advanced domain adaptation techniques, automated feature selection and representation learning approaches to develop intrusion detection systems that maintain resilience across diverse and evolving network environments.







\bibliographystyle{apalike}
{\small
\bibliography{example}}

\end{document}